\documentstyle[12pt]{article}
\newcommand{\be}{\begin{equation}}
\newcommand{\ee}{\end{equation}}
\newcommand{\bea}{\begin{eqnarray}}
\newcommand{\ena}{\end{eqnarray}}
\newcommand{\beano}{\begin{eqnarray*}}
\newcommand{\enano}{\end{eqnarray*}}
\newcommand{\sect}[1]{\setcounter{equation}{0}\section{#1}}
\newcommand{\vs}[1]{\rule[- #1 mm]{0mm}{#1 mm}}
\newcommand{\hs}[1]{\hspace{#1 mm}}
\newcommand{\frab}[2]{\frac{\displaystyle{#1}}{\displaystyle{#2}}}

\newcommand{\half}{\frac{1}{2}}

\newcommand{\alp}{\alpha}

\newcommand{\var}{\varphi}
\newcommand{\ca}{\mbox{$\cal{A}$}}
\newcommand{\cb}{\mbox{$\cal{B}$}}

\newcommand{{\cg}}{\mbox{$\cal{G}$}}

\newcommand{\cq}{\mbox{$\cal{Q}$}}
\newcommand{\cs}{\mbox{$\cal{S}$}}

\newcommand{\cw}{\mbox{$\cal{W}$}}
\newcommand{\prt}{\partial}
\newcommand{\eps}{\epsilon}
\newcommand{\und}[1]{\underline{#1}}

\newcommand{\r}[1]{{\cal R}_{#1}}
\newcommand{\rF}[1]{{\cal R}^F_{#1}}
\newcommand{\rB}[1]{{\cal R}^B_{#1}}
\newcommand{\rpi}[1]{{\cal R}_{#1}^{\pi}}
\newcommand{\Fd}{\mbox{Fund$^{\mbox{le}}$}}
\newcommand{\wh}[1]{\widehat{#1}}
\newcommand{\normal}[1]{{\normalsize{#1}}}
\newcommand{\mb}[1]{\hs{5}\mbox{#1}\hs{5}}

\newcommand{\0}[1]{\mbox{$\stackrel{\mbox{\scriptsize o}}{#1}$}}
\newcommand{\jok}{\mbox{\scriptsize{JOK}}}
\newcommand{\nn}{\mbox{\Large{$n$}}}

\newcommand{\NP}[1]{Nucl.\ Phys.\ {\bf #1}}
\newcommand{\PL}[1]{Phys.\ Lett.\ {\bf #1}}

\newcommand{\CMP}[1]{Comm.\ Math.\ Phys.\ {\bf #1}}

\newcommand{\IJMP}[1]{Int. Journ.\ Mod.\ Phys.\ {\bf #1}}

\newcommand{\ncon}[1]{\raisebox{-7mm}[1ex][-7mm]
                      {$\stackrel{\displaystyle #1}{\con}$}}
\newcommand{\con}{\rule[4mm]{0.2mm}{1.5mm}\rule[4mm]{1cm}{0.2mm}
                  \rule[4mm]{0.2mm}{1.5mm}}
\newcommand{\lcon}[1]{\raisebox{-7mm}[1ex][-7mm]
                      {$\stackrel{\displaystyle #1}{\hspace{-8mm}\noc}$}}
\newcommand{\noc}{\rule[4mm]{0.2mm}{1.5mm}\rule[4mm]{1.8cm}{0.2mm}
                  \rule[4mm]{0.2mm}{1.5mm}}

\begin{document}
\renewcommand{\thefootnote}{\fnsymbol{footnote}}
\newpage
\pagestyle{empty}
\setcounter{page}{0}
\rightline{May 93}
\rightline{NORDITA 93/39 P}
\vs{20}

\begin{center}

{\LARGE {\bf $OSp(1|2)$ and $Sl(2)$ reductions in
generalised super-Toda models
and factorization of spin 1/2 fields.}}\\[2cm]

{\large E. Ragoucy\footnote{On leave from absence of
Laboratoire de Physique Th\'eorique
{\small E}{\normal N}{\large S}{\Large L}{\large A}{\normal P}{\small P}
Chemin de Bellevue BP 110, F - 74941 Annecy-le-Vieux Cedex, France}}\\[.4cm]
{\em{ NORDITA}}
\\
{\em Blegdamsvej 17, DK-2100 Copenhagen \O, Denmark}\\[.5cm]

\end{center}
\vs{20}

\centerline{\bf Abstract}

\indent

I show that the classical Toda models built on superalgebras, and
obtained from a reduction with respect to an $Sl(2)$ algebra, are
"linearly supersymmetrizable" (by adding spin 1/2 fields) if and only if the
$Sl(2)$ is the bosonic part of an $OSp(1|2)$ algebra. In that case, the
model is equivalent to the one constructed from a reduction with
respect to the $OSp(1|2)$ algebra, up to spin 1/2 fields. The
corresponding $W$ algebras
are related through a factorization of spin 1/2 fields (bosons and
fermions). I illustrate this factorization on an example: the
superconformal algebra built on $Sl(n+1|2)$.
\vs{20}

\newpage
\pagestyle{plain}

\renewcommand{\thefootnote}{\arabic{footnote}}
\setcounter{footnote}{0}
\newpage

\sect{Introduction}

\indent

In the search of classifying all the $W$ (super)algebras \cite{BS}, the
generalised (super) Toda models play a central role \cite{ORaf,Sav}.
Indeed, for the study of $W$ algebras, two
important ways have been distinguished:
the direct computation with the use of Jacobi
identities \cite{klaus,Kausch}, or the construction of the $W$
algebras as symmetry of the Toda models.
In these models,
the $W$ algebra appears as a symmetry of the Toda action, and the
formalism of hamiltonian reduction provides a natural framework to
study their Operator Product Expansion (OPE), as well as their free
fields
realization. Although it has been shown \cite{Kausch,Bonn}
that (at least at the quantum level) there are some
$W$ algebras that do not seem to be obtained from Toda models, the
classification of the "Toda-$W$ algebras" is
nevertheless a first step in the classification of all the $W$
algebras. Progress have been made in that direction, specially in the
classical case, where all the different Hamiltonian reductions leading
to $W$ algebras have been classified
\cite{Oraf.class}, as well as the (super)spin
contents of the corresponding $W$ algebra\footnote{The spin contents for $W$
algebras built on $Sl(2)$ reduction of superalgebras was not
explicitly given in \cite{nous}. There was nevertheless all the tools
needed for such a classification, and I develop the technic
in this paper} \cite{nous}. The Toda-$W$ algebras are classified  by
the embedding of $Sl(2)$ in Lie (super)algebras \ca, and the
superconformal Toda-$W$ algebras by the $OSp(1|2)$ subalgebras of Lie
superalgebras. Due to the characterization of $OSp(1|2)$ and $Sl(2)$
subalgebras, it means that one can can exhaustively label the $W$
algebras as \cw(\ca, \cs) where \cs\ is a sub(super)algebra of \ca\
that defines the $Sl(2)$ or $OSp(1|2)$ embeddings (see for instance
\cite{nous} for details).
However, if this classification of Toda-$W$ algebras is exhaustive, it
is still unknown whether the $W$ algebras are all distinct, and if they are
some inclusions (or weaker relations) between them. In this paper, I
will show how to relate some of the $W$ algebras obtained from
$Sl(2)$ reduction
to the ones we get from $OSp(1|2)$ reductions. I will also show that the
$OSp(1|2)$ reduction is the only way that provides a superconformal
(i.e. linearly supersymmetric)
$W$ algebra in the Toda models framework.

\indent

This paper is organised as follow. In section \ref{osp12}, I recall
the basic tools that are needed to build a generalised superconformal
Toda model from a Wess-Zumino-Witten (WZW) action.
In particular, the role of the $OSp(1|2)$ subalgebra (of
the Lie superalgebra the model is built on) is emphasized.
Then, the section \ref{sl2} presents the case of
generalized Toda models associated to $Sl(2)$ reduction of Lie
superalgebras and show that the models which can be linearly
supersymmetrized are equivalent to the models presented in section
\ref{osp12}. In particular, it is shown that there exists a gauge
fixing where the actions differ only by kinetic terms in spin 1/2
fields. As a consequence, the corresponding $W$ algebras (symmetries
of the actions) are equivalent through a factorization of the spin 1/2
fields. These points are enlightened by an example in the section
\ref{sln2}: the quadratic superconformal algebra obtained from
the $Sl(n+1|2)$
superalgebra. I compute the OPEs and the free fields realization
in an $N=1$ formalism, then factorize the fermions of the algebra
and show that we recover the quadratic superconformal algebra obtained from the
reduction with respect to $Sl(2)$, that was already studied in
\cite{JOK}. Finally, I list in the conclusion the questions that are
still opened.

\section{$OSp(1|2)$ reductions \label{osp12}}

\indent

Let us first recall the basic ingredients that one needs to make a
constrained super-WZW model \cite{supAb}. One starts with a supersymmetric WZW
model based on a Lie superalgebra \ca. Its action reads
\bea
S_0(G) &=& \frac{\kappa}{2}\int d^2xd^2\theta
< G^{-1}D_+ \hat{G}, \hat{G}^{-1}D_-G> + \\
&& +
\int d^2xd^2\theta dt <G^{-1}\prt_tG, (G^{-1}D_+\hat{G})(\wh{G}^{-1}D_-G)
+(G^{-1}D_-\hat{G})(\wh{G}^{-1}D_+G)>\nonumber
\ena
where $<\ ,\ >$ is the non-degenerate scalar product on \ca, and
$D_\pm$ are the fermionic derivatives.
$G(x,\theta)$ is a superfield that belongs to the supergroup \cg\
associated to \ca: let me recall that when writing
$G(x,\theta)=expV(x,z)$, we will have $V(x,\theta)=V^a(x,\theta)t_a$
with $t_a$ a basis of \ca\ and $V^a(x,\theta)$ a bosonic (fermionic)
superfield when $t_a$ is a commuting (anti-commuting) generator.
I have also introduced the isomorphism $\wh{\ }$ of \ca\ that changes
the sign in front of the fermionic superfields. The action of
$\wh{\ }$ is also trivially extended to \cg. I recall that I consider
only the classical case. $S_0$ is invariant under the
transformations
\be
G(x_+,x_-,\theta_+,\theta_-) \rightarrow
L(x_-,\theta_-) G(x_+,x_-,\theta_+,\theta_-) R(x_+,\theta_+)
\mb{with} L,R\in\cg
\ee
The corresponding currents $J=\kappa D_-GG^{-1}$ and
$\bar{J}=\kappa \hat{G}^{-1}D_+G$ form two copies of the super
Kac-Moody (KM)
algebra based on \ca. Note that $J$ and $\bar{J}$ are fermionic superfields.
The equations of motion for $G$ are just the currents conservation laws:
\be
D_-J=0 \mb{and} D_+\bar{J}=0
\ee
$S_0(G)$ is also invariant under superconformal transformations,
generated by
\be
T_0=\frac{1}{3\kappa^2}Str(J\hat{J}J)+\frac{1}{2\kappa}Str(JD_-J)
 \mb{and} \bar{T_0}=\frac{1}{3\kappa^2}
Str(\bar{J}\bar{J}\hs{-4}\hat{\hs{4}\bar{J}})
+\frac{1}{2\kappa}Str(\bar{J}D_+\bar{J})
\ee
The superconformal properties of the currents are encoded in the
superOPEs\footnote{Stricly
speaking, I should write Poisson Brackets instead
of OPEs. It is nevertheless possible to use the OPEs formalism with the
correspondence $\delta(x-x')\equiv\frac{1}{x-x'}$. Note however that these
"classical" OPEs obey to the Leibniz rule, whereas the quantum ones do not.}.
\beano
\ncon{T(Z_-)J(X_-)} &=& \half\frac{\eta_--\theta_-}{(Z_--X_-)^2}J(X_-)
+\frac{1/2}{Z_--X_-}D_-J(X_-)
+\frac{\eta_--\theta_-}{Z_--X_-} D^2_-J(X_-) \nonumber\\
\ncon{\bar{T}(Z_+)\bar{J}(X_+)} &=& \half\frac{\eta_+-\theta_+}{(Z_+-X_+)^2}
\bar{J}(X_+) +\frac{1/2}{Z_+-X_+}D_-\bar{J}(X_+)
+\frac{\eta_+-\theta_+}{Z_+-X_+} D^2_-\bar{J}(X_+) \nonumber\\
\ncon{T(Z_-)\bar{J}(X_+)} &=& 0 \nonumber\\
\ncon{\bar{T}(Z_+)J(X_-)} &=& 0 \nonumber
\enano
with $Z_\pm=(z_\pm,\eta_\pm)$, $X_\pm=(x_\pm,\theta_\pm)$
and $Z-X=z-x-\eta\theta$. These OPEs
show that $J$ (resp. $\bar{J}$) is a primary superfield of dimensions
$(\half,0)$ (resp. $(0,\half)$).

\indent

To define a constrained WZW models, one first
introduces a grading operator $H$ of \ca:
\be
\ca=\oplus_{j\in\half{\bf Z}} \ca_j, \mb{with} [H,X_j]=jX_j\ \ \
\forall X_j\in\ca_j
\ee
For the model to be well-defined \cite{ORaf},
I will choose $H$ to be the Cartan
operator of some $OSp(1|2)$ subsuperalgebra of \ca. I will call
\be
\ca_+=\oplus_{j>0}\ca_j \mb{and} \ca_-=\oplus_{j<0}\ca_j
\ee
and $\cg_{\pm,0}$ the subgroups associated to $\ca_{\pm,0}$.

\indent

Then, one wants to gauge the left (right) action of $\cg_+$ ($\cg_-$).
For that aim, one makes the transformation
\be
G\ \longrightarrow\ \alp_+G\alp_- \mb{where} \alp_{\pm}\in\cg_{\pm}
\ee
Using the Polyakov-Wiegmann relation
\be
S_0(G_1G_2)=S_0(G_1)+S_0(G_2)+\kappa\int d^2xd^2\theta
<G_1^{-1}D_+\hat{G_1},(D_-G_2)G_2^{-1}>
\ee
we are led to
\bea
S_0(\alp_+G\alp_-) \,=\, S_0(G)+ \kappa\int &d^2xd^2\theta& \left\{
<A_+,(D_-G)G^{-1}>+ <G^{-1}D_+\hat{G},A_->+ \right.\nonumber\\
&&\ \left.+<G^{-1}A_+\hat{G},A_-> \right\}
\ena
with the gauge superfields
\be
A_+=\alp_+^{-1}D_+\hat{\alp}_+\in\ca_+ \mb{and}
A_-=(D_-\alp_-)\alp_-^{-1}\in\ca_-
\ee
Finally, denoting by $F_{\pm}$ the fermionic roots of the $OSp(1|2)$
subsuperalgebra whose Cartan generator is $H$, the complete gauge-invariant
action which leads to the non-Abelian superconformal Toda models reads:
\bea
S(G,A_+,A_-) \,=\, S_0(G)+ \kappa\int &d^2xd^2\theta& \left\{
<A_+,(D_-G)G^{-1}-F_->+ <G^{-1}D_+\hat{G}-F_+,A_-> \right.\nonumber\\
&&\ \left. +<G^{-1}A_+\hat{G},A_-> \right\} \label{Stod}
\ena
This action is invariant under the superconformal transformations generated
by
\be
T_H(X)=T_0(X)+<H,D^2_-J(X)> \mb{and}
\bar{T}_H(X)=\bar{T}_0(X)-<H,D^2_+\bar{J}(X)>
\label{eqTH}
\ee
Note that because of the new term in $T_H$, the different components of $J$
will have a superspin\footnote{I remind that a (chiral) superfield of
superspin $s$ is constructed on one field of spin $s$ and one field of spin
$s+\half$.} which depends on their grade under $H$:
\beano
J(X_-) &=& \sum_j J_j(X_-)t_j \mb{with}
\mbox{superspin}(J_j)=(\half+j,0) \mb{if} t_j\in\ca_j \\
\bar{J}(X_+) &=& \sum_j \bar{J}_j(X_+)t_j \mb{with}
\mbox{superspin}(\bar{J}_j)=(0,\half-j) \mb{if} t_j\in\ca_j
\enano
In particular, the $J$-components (resp. $\bar{J}$-components)
of negative (resp. positive) grades will have a negative spin,
so that they should be set to zero. This is exactly what is required by the
equations of motion of $A_+$ and $A_-$:
\[
J|_{\ca_-}=F_- \mb{and} \bar{J}|_{\ca_+}=F_+
\]
Using a Gauss decomposition
\be
G=G_+G_0G_- \mb{with} G_\pm\in\cg_\pm,\ G_0\in\cg_0
\ee
these constraints can be set in a nice way:
\beano
&& (D_-G_-)G^{-1}_- +\hat{G}_-A_-G_-^{-1} =\hat{G_0}^{-1}F_-G_0
\nonumber \\
&& G^{-1}_+D_+\hat{G}_+ +G_+^{-1}A_+\hat{G}_+ =G_0F_+\hat{G}^{-1}_0
\enano
The action is of course invariant under the gauge transformations:
\bea
G &\rightarrow& L_+GR_- \mb{with} L_+(X_+,X_-)\in\cg_+\ ,\
R_-(X_+,X_-)\in\cg_- \nonumber \\
A_+ &\rightarrow& L_+A_+\hat{L}_+^{-1}-(D_+\hat{L}_+)\hat{L}_+^{-1}
\label{gtransfo}\\
A_- &\rightarrow& \hat{R}^{-1}_-A_-R_- -\hat{R}_-^{-1}D_-R_-\nonumber
\ena
One can use this gauge invariance to set $A_+=0$ and $A_-=0$. This partially
fixes the gauge freedom.
With the residual gauge transformations (associated to $L_+(X_+)$ and
$R_-(X_-)$) two gauge fixing are in general used:

In the {\bf diagonal gauge}, the currents read
\be
\begin{array}{lll}
J_{diag.}(x_-,\theta_-) &=\, F_- + \sum_i \Phi^i(x_-,\theta_-)t_i & \\
&& \ \mbox{ with }t_i\mbox{ basis of }{\cb}_0 \\
\bar{J}_{diag.}(x_+,\theta_+) &=\, F_+ +
\sum_i \bar{\Phi}^i(x_+,\theta_+) t_i&
\end{array}
\ee
where ${\cb}_0=\ca_0$ if $H$ is such that all the (fermionic)
bosonic superfields have (half) integer superspin. If it is not the
case, ${\cb}_0=\ca_0\oplus\cq_{\half}\oplus\cq_{-\half}$, with
$\cq_{\pm\half}$ subspaces of $\ca_{\pm\half}$ determined through a
halving (see \cite{ORaf}).

The {\bf highest weights, or Drinfeld-Sokolov (DS) gauge}
\cite{DS}:
\bea
J_{DS}(x_-,\theta_-) &=& F_- + \sum_a \cw^a(x_-,\theta_-) e_a\
\mbox{ with } [F_+,e_a\}=0\\
\bar{J}_{DS}(x_+,\theta_+) &=& F_+ + \sum_a \bar{\cw}^a(x_+,\theta_+) e'_a\
\mbox{ with } [F_-,e'_a\}=0
\ena

\indent

Let us look at the superconformal properties in each of these gauges.
The $\Phi$'s and $\bar{\Phi}$'s are superfields of superspin
$(\half,0)$ and $(0,\half)$ respectively (because the $t_i$'s belong
to $\ca_0$).
They are all primary, but the one associated to $H$ (because to the
scalar product with $H$ in \ref{eqTH}).
The $\cw$'s (and $\bar{\cw}$'s) are all primary superfields, the superspin
of which being given by the
$OSp(1|2)$ decomposition of the adjoint representation of \ca \cite{nous}.
One can show that the $\cw$ superfields are gauge invariant
polynomials \cite{ORaf}, so that the Poisson
bracket of two $\cw$ generators closes (in the enveloping algebra of
the $\cw$'s): they form the superconformal $\cw$ algebra.
Going from one
gauge to the other allows us to have a free super-fields realization
of the super $\cw$ generators (super-Miura transformation).
All these points will be developed on an example in the section \ref{sln2}.

\section{$Sl(2)$ reduction \label{sl2}}

\indent

Instead of using a superfield formalism, one can do the previous
work with a field formalism, as for WZW models built on Lie algebras.
Since we will deal with
fields, the model will not be explicitly supersymmetric. To be
precise, if I use "superconformal" to distinguish the "linearly
supersymmetric" from the "non-linearly supersymmetric", the model
will be conformally but not superconformally
invariant. Thus, the super Virasoro algebra will not
be a subalgebra of the $W$ algebra under consideration. As the
superconformal algebra is associated with $OSp(1|2)$ algebra, whereas
the conformal one relies on the $Sl(2)$ algebra, it is easy to deduce
that the reduction and the constraints will be
associated with an $Sl(2)$ embedding in \ca.
Of course, there will be strong
connections between the two approaches: it has been already
shown\cite{supAb} that
in the Abelian case the "non superconformal" model can be obtained from
the superconformal one by choosing a gauge for the supersymmetry.
I will show that
this property extends to any of the $Sl(2)$ subalgebras that are the
bosonic part of a $OSp(1|2)$ superalgebra.

\subsection{Generalised conformal Toda action}

\indent

To construct a generalised Toda action with a $Sl(2)$ reduction, we
follow the same steps as in the previous section, replacing
superfields by fields. I briefly recall the main steps.

We start with the WZW action built on the superalgebra \ca:
\be
S_0(g) \,=\, \frac{\kappa}{2}\int d^2x <g^{-1}\prt_+g,g^{-1}\prt_-g>
+\int d^2xdt <g^{-1}\prt_tg,[g^{-1}\prt_+g,g^{-1}\prt_-g]>
\ee
$S_0(g)$ is invariant under the KM transformations generated by
$j(x_-)=\kappa (\prt_-g)g^{-1}$
and $\bar{\jmath}(x_+)=\kappa g^{-1}\prt_+g$.
It is also conformally invariant with respect to
$t_0(x_-)=\frac{1}{2\kappa}tr(j\prt_-j)$ and
$\bar{t}_0(x_+)=\frac{1}{2\kappa}tr(\bar{\jmath}\prt_+\bar{\jmath})$.

Then, we choose a Cartan generator $H$ of an $Sl(2)$ subalgebra of
\ca, and gauge the (left) right action of ($\cg_+$) $\cg_-$. Then, denoting
by $E_\pm$ the roots of the $Sl(2)$ subalgebra under consideration,
the generalised conformal Toda action built on the superalgebra \ca\ reads:
\be
S(g,a_+,a_-)= S_0(g) -2\int d^2x\left\{<a_-,g^{-1}\prt_+g+E_->
+<(\prt_-g)g^{-1}-E_+,a_+> +<g^{-1}a_+g,a_-> \right\}
\label{Snonsup} \\
\ee
The conformal generators are
\be
t_H(x)=t_0(x)+<H,\prt_-j(x)> \mb{and}
\bar{t}_H(x)=\bar{t}_0(x)-<H,\prt_+\bar{\jmath}>
\ee
Note that the spin of the components of $j$ and $\bar{\jmath}$ are
\beano
j(x_-) &=& \sum_\ell j_\ell(x_-)t_\ell \mb{with}
\mbox{spin}(j_\ell)=(1+\ell,0) \mb{if} t_\ell\in\ca_\ell \\
\bar{\jmath}(x_+) &=& \sum_\ell \bar{\jmath}_\ell(x_+)t_\ell \mb{with}
\mbox{spin}(\bar{\jmath}_\ell)=(0,1-\ell) \mb{if} t_\ell\in\ca_\ell
\enano
The $W$ generators are defined in the DS gauge as:
\beano
j(x_-) &=& E_-+ \sum_a W^a(x_-)e_a \mb{with} [E_+,e_a]=0 \\
\bar{\jmath}(x_-) &=& E_++ \sum_a \bar{W}^a(x_-)e'_a \mb{with} [E_-,e'_a]=0
\enano
and the spin contents is given by the $Sl(2)$ decomposition of \ca.

\subsection{Wess-Zumino gauge for the supersymmetry \label{sl2.gaug}}

\indent

Now, I want to show how to relate the action (\ref{Snonsup}) to the action
(\ref{Stod}).
As the calculation is the
same as in the Abelian case, I will just sketch the proof: for
details, see \cite{supAb}.
I come back to the action (\ref{Stod}). From the gauge transformation
laws (\ref{gtransfo}),
it is easy to see that we can choose a gauge such that:
\bea
&& A_+|_{\theta=0}=0 \mb{and} (D_+A_+)|_{\theta=0}=0 \label{c1}\\
&& A_-|_{\theta=0}=0 \mb{and} (D_-A_-)|_{\theta=0}=0 \label{c2}\\
&& J|_{\theta=0}\in(\ca_-\oplus\ca_0)
\mb{and} \bar{J}|_{\theta=0}\in(\ca_+\oplus\ca_0) \label{c3}
\ena
The fields $(D_+D_-A_+)|_{\theta=0}$ and $(D_+D_-A_-)|_{\theta=0}$ are
auxiliary fields for the supersymmetry: their equation of motion
together with (\ref{c3}) show that the restriction of $J|_{\theta=0}$ (resp
$\bar{J}|_{\theta=0}$) on $\ca_+$ (resp. $\ca_-$) is equal to $F_-$
(resp. $F_+$). Denoting $J|_{\theta=0}=F_-+\chi_-$ and
$\bar{J}|_{\theta=0}=F_++\chi_+$ where $\chi_\pm$ are some fields of
grade 0, one can write the component form of the action:
\bea
&&S \,=\, \frac{\kappa}{2}\int d^2x <g^{-1}\prt_+g,g^{-1}\prt_-g>
+\int d^2xdt <g^{-1}\prt_tg,[g^{-1}\prt_+g,g^{-1}\prt_-g]>+
\label{Scompo} \\
&&\ -2\int d^2x\left\{<a_-,g^{-1}\prt_+g+F_-^2>
+<(\prt_-g)g^{-1}-F_+^2,a_+> +<g^{-1}a_+g,a_-> \right\}+
\nonumber \\
&&\ +\int d^2x\left\{ <\chi_-,\prt_+\chi_-> +<\chi_+,\prt_-\chi_+>
-2<a_+,[\chi_-,F_+]> +2<[\chi_+,F_-],a_-> \right\} \nonumber
\ena
where $g(x)=G(X)|_{\theta=0}$ is an element of the group built on \ca,
and $a_+=(D_+\hat{A}_+)|_{\theta=0}$,
$a_-=(D_-A_-)|_{\theta=0}$ are the physical gauge fields for the
"non superconformal" action. Their equation of motions lead to
constraints on the (non supersymmetric) currents
$j(x)=(D_-J)|_{\theta=0}=(\prt_-g)g^{-1}$ and
$\bar{\jmath}(x)=(D_+\bar{J})|_{\theta=0}=g^{-1}\prt_+g$ :
\bea
j(x)|_{\ca_-} &=& \half\{F_-,F_-\} -[\chi_-,F_-]-ga_-g^{-1}
=E_--[\chi_-,F_-]-ga_-g^{-1} \\
\bar{\jmath}(x)|_{\ca_+} &=& \half\{F_+,F_+\} +[\chi_+,F_+]-g^{-1}a_+g
=E_++[\chi_+,F_+]-g^{-1}a_+g
\ena
where $E_\pm$ are the (bosonic) roots of the $Sl(2)$ subalgebra
whose Cartan generator is $H$. One recovers the usual constraints of a
"non-superconformal" generalised Toda model\footnote{I denote by
"non-superconformal" the actions that do not possess any \und{linear}
supersymmetry invariance. The
action \ref{Scompo} \und{is} (non linearly) supersymmetric, but the
supersymmetry transformations have quadratic terms, as it was shown in
\cite{supAb}}
implemented by the
spin-$\half$ fields that appear in the halving procedure \cite{ORaf}.

\indent

Note that because $\chi_\pm$ belong to $\ca_0$, they will have a spin
$\half$, but they may be bosons, depending on the characteristic
(commuting or anti-commuting) of the \ca-generators they are carried by.
As far as spin 1/2 fermions are concerned, it is known \cite{fact}
that the corresponding \cw\ algebras can be factorized. However, for
spin 1/2 bosons, although nothing has been proved until now, it seems
that the factorization works as well.

\indent

Let us now fix the gauge $a_+=0=a_-$. Then,
\bea
S &=& \frac{\kappa}{2}\int d^2x <g^{-1}\prt_+g,g^{-1}\prt_-g>
+\int d^2xdt <g^{-1}\prt_tg,[g^{-1}\prt_+g,g^{-1}\prt_-g]>
\nonumber\\
&& +\int d^2x\left\{ <\chi_-,\prt_+\chi_-> +<\chi_+,\prt_-\chi_+>
\right\} \nonumber
\ena
the component action is the sum of a usual ("non-superconformal")
generalised Toda
model built on \ca\ (in the gauge $A=0$) and of the action of free
spin $\half$ fields. This
means that the $W$ algebra which is a symmetry of this generalised Toda model
can be deduced from the superconformal $\cw$ algebra constructed from
the action (\ref{Stod}) by factorizing out the spin $\half$ fields (bosons
and fermions)
$\chi_\pm$. I will show below that
the converse is true: when starting from a $W$ algebra (obtained from
the $Sl(2)$ reduction of \ca), if one can make it superconformal by
adding spin-$\half$ fields, then the $Sl(2)$ subalgebra is the
bosonic part of an $OSp(1|2)$ sub-superalgebra of \ca. In that case, the
superconformal version of the $W$ algebra is obtained by the
$OSp(1|2)$ reduction of \ca.

\subsection{$Sl(2)$ decomposition of superalgebras and
"non-superconformal" $W$ algebras \label{sl2.nonsup}}

\indent

The way of decomposing a
superalgebra \ca\ with respect to $Sl(2)$ subalgebras is similar to
the $Sl(2)$ decomposition of
fundamental and adjoint representations of Lie algebra
(see \cite{nous} for details).
One first decomposes (w.r.t $Sl(2)$) the fundamental
representation of each simple Lie algebra entering in the bosonic part
of \ca. Then, the anticommuting generators being in their fundamental
representations ($(m,\bar{n})+(\bar{m},n)$ for $Sl(m|n)$, $(m,n)$ for
$OSp(m|n)$, and so on..), one get their $Sl(2)$ decomposition by a
simple product of $Sl(2)$ representations. Finally, the bosonic part
decomposition is obtained as in the Lie algebras case.
A good check of the results given in the
previous section is to compare the $Sl(2)$ and the $OSp(1|2)$
decompositions when the $Sl(2)$ is the bosonic part of $OSp(1|2)$.

\indent

Let us study an example in the $Sl(m|n)$ case.
I start with the principal embedding of
$Sl(2)$ in $Sl(p+1)\oplus Sl(p)$, considered as a subalgebra of
$Sl(m|n)$. We have the decompositions:
\beano
&&(m,\bar{n})+(\bar{m},n) \,=\, 2\left( D_{p/2}\oplus (m-p-1)D_0\right)
\times\left( D_{(p-1)/2}\oplus (n-p)D_0\right) \\
&& \ \ \,=\, 2\oplus_{j=1/2}^{p-1/2} D_j \oplus
2(n-q-1) D_{p/2} \oplus 2(m-p-1) D_{(p-1)/2}
\oplus 2(m-p-1)(n-p-2) D_0 \nonumber \\
&& Sl(m)\oplus Sl(n)\oplus U(1) \,=\,
\left(D_{p/2}\oplus(m-p-1)D_0\right)^2 \oplus \left(D_{(p-1)/2}\oplus
(n-p)D_0\right)^2 \ominus D_0\\
&&\ \ 2\oplus_{j=0}^{p-1} D_j \oplus
D_p \oplus 2(m-p-1) D_{p/2} \oplus 2(n-p-2) D_{(p-1)/2} \oplus \nonumber \\
&& \ \ \oplus\left[ (m-p-1)^2+(n-p-2)^2-1\right] D_0
\enano
which can be transcribed into fields contents:
\bea
\mbox{Bosonic fields} &:& W_{p+1}\ ;\ 2W_j \ (j=1,..,p)\
;\ 2(n-p)\, W_{(p+1)/2}\ ;\ 2(m-p-1)\, W_{p/2+1}
\nonumber \\
&& \left[ (m-p-1)^2+ (n-p)^2 -1\right] W_{1}\\
\mbox{Fermionic fields} &:& 2W_{j+1/2}\ (j=1,..,p)\ ;\ 2(m-p-1)\,
W_{(p+1)/2} \ ;\ 2(n-p)\, W_{p/2+1}\nonumber \\
&&  2(m-p-1)(n-p)\, W_1
\ena

Now, if one does the same work with an $OSp(1|2)$ subalgebra
which is the principal embedding of $Sl(p+1|p)\subset Sl(m|n)$, one gets:
\bea
\Fd &=& \r{p/2}\oplus(m-p-1)\r{0}\oplus(n-p)\rpi{0} \\
\mbox{Adj} &=& \oplus_{j=0}^{p-1} \left(\rB{j}\oplus\rF{j+\half}\right)
\oplus \rB{p}\oplus 2(m-p-1)\rF{p/2} \oplus
2(n-p)\rB{p/2} \oplus \nonumber\\
&& \oplus \left[ (m-p-1)^2+ (n-p)^2 -1\right] \rF{0} \oplus
2(m-p-1)(n-p) \rB{0}
\ena
where $\r{},\rpi{},\rF{},\rB{}$ are $OSp(1|2)$ representations, made
of two $Sl(2)$ representations: in the fundamental representation of
$Sl(m|n)$, $\r{j}=(D_j,D_{j-\half})$ and
$\rpi{j}=(D_{j-\half},D_j)$ where the first term in the pair belongs to
$Sl(m)$ representation while the second is in $Sl(n)$ one; in the
adjoint representation of $Sl(m|n)$,
$\rF{q}=(D_q,D_{q-\half})$ and $\rB{q}=(D_q,D_{q-\half})$ with $D_q$
built on anticommuting (resp. commuting) generators for $\rF{}$ (resp.
$\rB{}$). The adjoint decomposition gives rise to
the following superspin contents for the super $\cw$ algebra:
\bea
\mbox{Bosonic superfields} &:& \cw_j \ (j=1,..,p)\ ;\ 2(n-p)\,
\cw_{(p+1)/2} \nonumber \\
&& 2(m-p-1)(n-p)\, \cw_{1/2} \\
\mbox{Fermionic superfields} &:& \cw_{j+1/2}\ (j=0,..,p)\ ;\ 2(m-p-1)\,
\cw_{(p+1)/2} \nonumber \\
&& \left[ (m-p-1)^2+ (n-p)^2 -1\right] \cw_{1/2}
\ena
In terms of fields, one writes the super $\cw$ algebra as:
\bea
\mbox{Bosonic fields} &:& W_{p+1}\ ;\ 2W_j \ (j=1,..,p)\
;\ 2(n-p)\, W_{(p+1)/2}\ ;\ 2(m-p-1)\, W_{p/2+1}
\nonumber \\
&& 2(m-p-1)(n-p)\, W_{1/2} \ ;\
\left[ (m-p-1)^2+ (n-p)^2 -1\right] W_{1}\\
\mbox{Fermionic fields} &:& 2W_{j+1/2}\ (j=1,..,p)\ ;\ 2(m-p-1)\,
W_{(p+1)/2} \ ;\ 2(n-p)\, W_{p/2+1}\nonumber \\
&& \left[ (m-p-1)^2+ (n-p)^2 \right] W_{1/2} \ ;\
 2(m-p-1)(n-p)\, W_1
\ena
As announced, the two approaches give the same fields contents, except
for spin $\half$ fields that are absent from the $Sl(2)$ reduction.
This calculation can be done for all kind of $OSp(1|2)$
embeddings in $Sl(m|n)$, the result is the same. It extends for
$OSp(m|n)$ algebras by folding of $Sl(m|n)$ ones\footnote{One can also
do the calculation by hand}. Of course when the
$Sl(2)$ subalgebra is not the bosonic part of an $OSp(1|2)$
superalgebra, the corresponding $W$ will not be equivalent to one
obtained from an $OSp(1|2)$ algebra, but it will not possible to make
it
superconformal. In the table \ref{t1}, I give the basic $Sl(2)$
embeddings that furnish a "virtually superconformal" $W$ algebra, as well as
the $OSp(1|2)$ algebra which is in correspondence. Any
 "virtually superconformal" $W$ algebra will be obtained from a (or a sum
of) basic $Sl(2)$.

\begin{table}[h]
\begin{tabular}{|c|c|c|}
\hline
&& \\
\mbox{Superalgebra} & $Sl(2)$ \mbox{embedding}
& \mbox{Corresponding }$OSp(1|2)$\\
&& \\
\hline
&& \\
$Sl(n+1|m+1) $
& $Sl(p+1)\oplus Sl(p) \hs{7} p\leq(m,n-1)$ & $Sl(p+1|p)$ \\
& $Sl(p)\oplus Sl(p+1) \hs{7} p\leq(m-1,n)$ & $Sl(p|p+1)$ \\
&& \\
\hline
&& \\
$OSp(m|2n) $
&$SO(2p+1)\oplus Sp(2p) \hs{7} p\leq([(m-1)/2],n)$ & $OSp(2p+1|2p)$\\
&$SO(2p-1)\oplus Sp(2p) \hs{7} p\leq([(m+1)/2],n)$ & $OSp(2p-1|2p)$\\
&$SO(2p)\oplus Sp(2p) \hs{7} p\leq([m/2],n)$ & $OSp(2p|2p)$\\
&$SO(2p+2)\oplus Sp(2p) \hs{7} p\leq([(m-2)/2],n)$ & $OSp(2p+2|2p)$\\
& $Sl(p+1)\oplus Sl(p) \hs{7} p\leq([(m-2)/2],2n)$ & $Sl(p+1|p)$ \\
& $Sl(p)\oplus Sl(p+1) \hs{7} p\leq([m/2],2n-1)$ & $Sl(p|p+1)$ \\
&& \\
\hline
&& \\
$G(3)$
& $Sl(2)$ & $Sl(2|1)$ \\
& $Sl(2)'$ & $Sl(2|1)'$ \\
& $Sl(2)''$ & $OSp(1|2)$ \\
& $Sl(2)\oplus Sl(2)$ & $OSp(3|2)$ \\
& $Sl(2)\oplus Sl(2)\oplus Sl(2)$ & $D(2,1;3)$ \\
&& \\
\hline
&& \\
$F(4)$
& $Sl(2)$ & $Sl(2|1)$ \\
& $Sl(2)'$ & $Sl(1|2)$ \\
& $Sl(2)''$ & $OSp(2|2)$ \\
& $Sl(2)\oplus Sl(2)\oplus Sl(2)$ & $D(2,1;2)$ \\
&& \\
\hline
&& \\
$D(2,1;\alp)$
& $Sl(2)$ & $Sl(2|1)$ \\
& $Sl(2)'$ & $OSp(2|2)$ \\
& $Sl(2)\oplus Sl(2)\oplus Sl(2)$ & $D(2,1;\alp)$ \\
&& \\
\hline
\end{tabular}
\caption{Basic $Sl(2)$ embeddings that lead to a "virtually superconformal"
$W$ algebra \label{t1}}
\end{table}

\subsubsection{Quasi-Abelian $W$ algebras}

\indent

To end this section, I give a class of $Sl(2)$ reductions
that lead to "non-superconformal" $W$ algebras: the cases where the
$Sl(2)$ is the principal embedding in the bosonic part of \ca.
These models are very close to the Abelian Toda models in the sense
that $\ca_0$ is generated by the Cartan subalgebra only. However, they
are not Abelian, because one has to add some fermions of the $\ca_\pm$
spaces to build the model (halving).
Of
course, for the $W$ algebra to be "non-superconformal", \ca\ must
be a superalgebra with no principal $OSp(1|2)$ embedding.
That is, one must
choose \ca\ different from $Sl(m|n\pm1)$, $OSp(2m\pm1|2m)$,
$OSp(2m|2m)$, $OSp(2m+2|2m)$ and $D(2,1;\alp)$.

\indent

Let us start with $Sl(m+1|n+1)$ ($m\neq n\pm1$). The decomposition of the
fundamental of an $Sl(p+1)$ with respect to its principal embedding is
$\und{p}=D_{p/2}$, so that we have:
\bea
(m+1,\overline{n+1})+(\overline{m+1},n+1) &=& 2(D_{m/2}\times D_{n/2}) =
2\oplus_{j=\frac{|m-n|}{2}}^{(m+n)/2} D_j \nonumber\\
Sl(m+1)\oplus Sl(n+1)\oplus U(1) &=& (D_{m/2}\times D_{m/2})
\oplus (D_{n/2}\times D_{n/2})\ominus D_0 \nonumber \\
&=& \oplus_{j=1}^{m} D_j \oplus
\oplus_{j=1}^n D_j \oplus D_0
\ena

In the case of $OSp(2m+1|2n)$ algebras ($n\neq m,m-1$), we have
\bea
(2m+1,2n) &=& D_m\times D_{n-1/2} = \oplus_{j=|m-n+1/2|}^{m+n-1/2} D_j
\nonumber \\
SO(2m+1)\oplus Sp(2n) &=& \left.(D_m\times D_m)\right|_A \oplus
\left.(D_{n-1/2}\times D_{n-1/2})\right|_S \nonumber\\
&=& \oplus_{j=0}^{m-1} D_{2j+1} \oplus_{j=0}^{j=n-1} D_{2j+1} \label{Wb}
\ena
Let us remark that the well-known $WB(0,n)$ \cite{WB}
is the subcase $m=0$ in (\ref{Wb}).

For $OSp(2m|2n)$ algebras, the decompositions read
\bea
(2m,2n) &=& (D_{m-1}\oplus D_0)\times D_{n-1/2} = D_{n-1/2}
\oplus_{j=|m-n-1/2|}^{m+n-3/2} D_j \nonumber \\
SO(2m)\oplus Sp(2n) &=& \left.(D_{m-1}\oplus D_0)\times (D_{m-1}\oplus
D_0)\right|_A \oplus
\left.(D_{n-1/2}\times D_{n_1/2})\right|_S \nonumber\\
&=& D_{m-1} \oplus_{j=0}^{m-2} D_{2j+1}
\oplus_{j=0}^{j=n-1} D_{2j+1}
\ena

The calculation for $D(2,1;\alp)$ is the same as for $OSp(4|2)$: although
they are different, the
$W$ algebras have the same superspin contents.
For $G(3)$, the decomposition under the principal $Sl(2)$ of
$G_2\oplus Sl(2)$ leads to
\beano
(7,2) &=& D_3\times D_{1/2} \,=\, D_{7/2}\oplus D_{5/2} \\
G_2\oplus Sl(2) &=& D_5 \oplus 2D_1
\enano

Finally, for $F(4)$, we get
\beano
(2,8) &=& D_{1/2}\times (D_3\oplus D_0) \,=\, D_{7/2}\oplus D_{5/2}
\oplus D_{1/2} \\
Sl(2)\oplus SO(7) &=& D_5 \oplus D_3 \oplus 2D_1
\enano

The corresponding spin contents are gathered
in the table \ref{t2}. Note that all these
algebras have "good" statistic (bosons and fermions with usual spins).

\begin{table}[h]
\begin{tabular}{|c|c|c|}
\hline
&& \\
\mbox{Superalgebra} & \mbox{Spin contents} & \mbox{Spin contents} \\
& \mbox{(Bosons)} & \mbox{(Fermions)} \\
&& \\
\hline
&& \\
$Sl(m+1|n+1)$ &
$\begin{array}{c}W_j,\ (j=2,..,m+1) \\W_1 \\W_j,\ (j=2,..,n+1)
\end{array}$ & $2W_j\
(j=\frac{|m-n|}{2}, ..,\frac{m+n+2}{2})$ \\
&& \\
\hline
&& \\
$OSp(2m+1|2n)$ &
$\begin{array}{c}W_{2j},\ (j=1,..,m) \\W_{2j},\ (j=1,..,n)
\end{array}$ & $W_j \
(j=\frac{|2m-2n+1|}{2}+1,..,\frac{2m+2n+1}{2})$ \\
&& \\
\hline
&& \\
$OSp(2m|2n)$ & $\begin{array}{c}W_{2j},\ (j=1,..,m-1)\,;\, W_{m}
\\W_{2j},\ (j=1,..,n)
\end{array}$ & $\begin{array}{c}W_j \ (j=\frac{|2m-2n-1|}{2}+1
,..,\frac{2m+2n-1}{2})\\ W_{n+1/2} \end{array}$ \\
&& \\
\hline
&& \\
$G(3)$ & $ W_{6}
\, ;\, 2W_{2}$
& $ W_{9/2} \, ;\, W_{7/2}$ \\
&& \\
\hline
&& \\
$F(4)$ & $ W_{6}\, ;\, W_4 \, ;\, 2W_{2}$
& $ W_{9/2} \, ;\, W_{7/2} \, ;\, W_{3/2}$ \\
&& \\
\hline
\end{tabular}
\caption{Quasi-Abelian $W$ algebras \label{t2}}
\end{table}

\section{Factorisation of fermions in $Sl(n+1|2)$ case \label{sln2}}

\indent

In this section I will show explicitly how the factorization of the
spin $\half$ fermions works on one basic example: the quadratic superconformal
algebra built on $Sl(n+1|2)$ (i.e. the \cw[$A(n,1),A(1,1)$] algebra). This
algebra has been already studied in \cite{JOK} from the point of view
of $Sl(2)$ reduction. I study the case of the $OSp(1|2)$ reduction.
I will
give a free superfields realization of the superconformal $\cw$ algebra,
and give its superOPEs. Then, I will factorize the spin $1/2$
fermions and recover the free field realization given in \cite{JOK} for
$W$ algebra obtained by a reduction with respect to $Sl(2)$.
To avoid boring repetitions, I will focus on the $J$ part, but the results
are valid also for $\bar{J}$. I will not mention the subscript $"-"$ anymore,
and speak of the (super) spin $s$ instead of $(s,0)$.

\subsection{Constraints for the $Sl(n+1|2)$ quadratic superconformal algebra}

\indent

The algebra corresponds to the second line of table \ref{t1}
with $m=1$ and $p=0$. Let us first look at the superspin and
spin contents. It is very similar to the example presented in section
\ref{sl2.nonsup}. The results is:
\bea
\mbox{Bosonic superfields} &:& 2(n-1)\, \cw_{1} \nonumber \\
\mbox{Fermionic superfields} &:& \cw_{3/2} \ ;\ (n-1)^2\, \cw_{1/2}
\ena
We see that we will have a $Sl(n)\oplus U(1)$ super KM algebra.
Looking at the results of \cite{JOK}, one may be surprised, since they
get an $Sl(n+1)\oplus U(1)$ KM algebra. The difference comes from the
superfields formalism: in this formalism, we have a super Virasoro
subalgebra, but the price to pay is an (apparent) diminution
of the KM part from
$Sl(n+1)$ to (super) $Sl(n)$. The other spin 1 fields have been used to build
supersymmetry partners of the spin 3/2 fields. Indeed,
in terms of fields, one writes the super $\cw$ algebra as:
\bea
\mbox{Bosonic fields} &:& W_{2}\ ;\ (n+1)^2\, W_{1}\\
\mbox{Fermionic fields} &:& 2n\, W_{3/2} \ ;\ n^2\, W_{1/2}
\ena
showing that we have the right number of fields to construct an
$Sl(n+1)\oplus U(1)$ KM algebra.

\indent

Now, we start with an element $J$ in the super KM algebra of
$Sl(n+1|2)$ and want to impose some constraints on it. The Cartan
element for the $OSp(1|2)$ algebra under consideration is
\be
H=\left(
\begin{array}{ccc|cc}
&&& 0 & 0 \\
&\mbox{\Large{0}}& & \vdots & \vdots \\
&&& 0 & 0 \\
\hline
0 & \cdots & 0 & \half & 0 \\
0 & \cdots & 0 & 0 & -\half
\end{array}
\right)
\ee
which leads to the grades\footnote{$g$ is define by
$g_{ij}=$grade$(E_{ij})$ with $E_{ij}$ the matrix basis
$(E_{ij})_{pq}=\delta_{ip}\delta_{jq}$.}
\be
g=\left(
\begin{array}{ccc|cc}
&&& -\half & \half \\
& \mbox{\Large{0}} && \vdots & \vdots \\
&&& -\half & \half \\
\hline
\half & \cdots & \half & 0 & 1 \\
-\half & \cdots & -\half & -1 & 0
\end{array}\right)
\ee
and the constrained current reads
\be
J_{constr.}=\left(
\begin{array}{cccc|cc}
&&&& 0 & * \\
&& \mbox{\LARGE{*}} && \vdots & \vdots \\
&&&& 0 & * \\
&&&& 1 & * \\
\hline
* & \cdots & \cdots & * & * & * \\
0 & \cdots & 0 & -1 & 0 & *
\end{array}\right)
\ee
The two gauges described in section \ref{osp12} are:
\[
J_{diag.}=\left(
\begin{array}{cccc|cc}
&&&& 0 & 0 \\
&\frac{\Phi_1+\Phi_2}{n+1}{\bf{ 1_{n+1}}}& +\mbox{{M}}&
& \vdots & \vdots \\
&&&& 0 & 0 \\
&&&& 1 & 0 \\
\hline
0 & \cdots & \cdots & 0 & \Phi_1 & 0 \\
0 & \cdots & 0 & -1  & 0 & \Phi_2
\end{array}\right)
\mb{with} M\in Sl(n+1)
\]
and
\[
J_{DS}=\left(
\begin{array}{cccc|cc}
&&&0& 0 & G_n \\
&\frac{U}{n}{\bf 1_n}& +\mbox{{N}} &\vdots& \vdots & \vdots \\
&& & 0 & 0 & G_1 \\
0 &\cdots & 0 & U & 1 & G_0 \\
\hline
\bar{G_n} & \cdots & \bar{G_1} & -G_0 & U & T \\
0 & \cdots & 0 & -1 & 0 & U
\end{array}\right)
\mb{with} N\in Sl(n)
\]
where the highest weights gauge is defined with respect to the root
\be
F_+=\left(
\begin{array}{cccc|cc}
&&&& 0 & 0 \\
&& \mbox{\Large{0}} && \vdots & \vdots \\
&&&& 0 & 0 \\
&&&& 0 & 1 \\
\hline
0 & \cdots & 0 & 1 & 0 & 0 \\
0 & \cdots & 0 & 0 & 0 & 0
\end{array}\right)
\ee

\subsection{SuperOPEs of the \cw\ algebra \label{supope}}

\indent

First, I compute the OPEs of the $\cw$ algebra thanks to a
supersymmetric version of the so-called soldering procedure
\cite{poly}. Starting with an element $\Lambda(X)$ in
the super KM algebra $Sl(n+1|2)$, we make an infinitesimal gauge
transformation of $J_{DS}(Z)$:
\be
\delta_\Lambda\,J_{DS}(Z) =[\Lambda,J_{DS}(Z)\}+kD\Lambda
\ee
and ask this transformation to preserve the gauge: this imposes some
relations between the matrix elements of $\Lambda$ and allows us to
determine some of the $\Lambda_{jk}$'s
in terms of the \cw\ generators and of the other $\Lambda_{jk}$'s.
Then, replacing
these parameters in the variation of $J_{DS}$, we can compute its
infinitesimal gauge transformation. Finally, from the relation
\be
\delta_\Lambda\,J_{DS}(Z) =\int dxd\theta
Str(\lcon{\hat{\Lambda}J_{DS})(X)J_{DS}(Z)}
\ee
we are able to deduce the super OPEs:
\bea
\ncon{T(Z)T(X)} &=& k^2\left\{\frac{3}{2}\frac{\eta-\theta}{(Z-X)^2}T(X)
+\frac{1/2}{Z-X}DT(X) +\frac{\eta-\theta}{Z-X} D^2T(X) \right\} +\nonumber\\
&& +\frac{k^3}{(Z-X)^3}
\label{eq:i}\\
\ncon{T(Z)G_0(X)} &=& k^2\left\{\frac{\eta-\theta}{(Z-X)^2}G_0(X)
+\frac{1/2}{Z-X}DG_0(X) +\frac{\eta-\theta}{Z-X} D^2G_0(X) \right\}\\
\ncon{T(Z)G_p(X)} &=& k^2\left\{\frac{\eta-\theta}{(Z-X)^2}G_p(X)
+\frac{1/2}{Z-X}DG_p(X) +\frac{\eta-\theta}{Z-X} D^2G_p(X)
\right\}+ \nonumber\\
&& +\frac{k/2}{Z-X}\left(\frac{n-1}{n}U{G}_p + (NG)_p\right)+ \\
&& +\frac{\eta-\theta}{Z-X} \left[ k\frac{n-1}{n}G_pDU
+k[(DN)G]_p+(N^2G)_p \right]
\nonumber\\
\ncon{T(Z)\bar{G}_p(X)} &=& k^2\left\{
\frac{\eta-\theta}{(Z-X)^2}\bar{G}_p(X)
+\frac{1/2}{Z-X}D\bar{G}_p(X) +\frac{\eta-\theta}{Z-X}
D^2\bar{G}_p(X)\right\} + \nonumber\\
&& -\frac{k/2}{Z-X}\left( (\bar{G}N)_p+\frac{n-1}{n}U\bar{G}_p
\right)+\\
&& +\frac{\eta-\theta}{Z-X} \left[ -k\frac{n-1}{n}\bar{G}_pDU
-k(\bar{G}DN)_p-(\bar{G}N^2)_p \right]
\nonumber\\
\ncon{T(Z)U(X)} &=& 0 \hs{42}
\ncon{T(Z)N_{pq}(X)} \,=\, 0 \\
\ncon{G_0(Z)G_0(X)} &=& \frac{k^2}{4}\left\{
2\frac{\eta-\theta}{Z-X} \frac{1}{k^2}T(X) +\frac{2k}{(Z-X)^2} \right\}
\\
\ncon{G_0(Z)G_p(X)} &=& \frac{k}{2}\left\{
\frac{2}{Z-X}G_p +\frac{\eta-\theta}{Z-X}DG_p \right\}
+\frac{1}{2}\frac{\eta-\theta}{Z-X}(WG)_p
\\
\ncon{G_0(Z)\bar{G}_p(X)} &=& -\frac{k}{2}\left\{ \frac{2}{Z-X}\bar{G}_p
+\frac{\eta-\theta}{Z-X}D\bar{G}_p \right\}
+\frac{1}{2}\frac{\eta-\theta}{Z-X}(\bar{G}W)_p
\\
\ncon{G_0(Z)U(X)} &=& 0 \hs{42}
\ncon{G_0(Z)N_{pq}(X)} \,=\, 0 \\
\ncon{U(Z)G_p(X)} &=& \frac{\eta-\theta}{Z-X} G_p(X) \hs{24}
\ncon{U(Z)\bar{G}_p(X)} \,=\, -\frac{\eta-\theta}{Z-X} \bar{G}_p(X) \\
\ncon{N_{pq}(Z)G_r(X)} &=& \frac{\eta-\theta}{Z-X} \left(
\delta_{qr}G_p(X)- \frac{1}{n}\delta_{pq}G_r(X) \right)
\\
\ncon{N_{pq}(Z)\bar{G}_r(X)} &=& -\frac{\eta-\theta}{Z-X} \left(
\delta_{pr}\bar{G}_q(X)- \frac{1}{n}\delta_{pq}\bar{G}_r(X) \right)
\\
\frac{1}{k^2}\ncon{G_p(Z)\bar{G}_q(X)} &=& \frac{k\delta_{pq}}{(Z-X)^2}
+\frac{\eta-\theta}{Z-X}\left\{ -\delta_{pq}T(X)+
\frac{1}{k^2}(W^3)_{pq} -\frac{1}{k}(WDW)_{pq} \right\}+ \nonumber\\
&& -\frac{2}{k}\frac{\eta-\theta}{Z-X} W_{pq}G_0(X)
-\frac{\delta_{pq}}{k}\left\{\frac{2}{Z-X} G_0(X)
+\frac{\eta-\theta}{Z-X}DG_0(X) \right\}+ \nonumber\\
&& - \frac{\eta-\theta}{(Z-X)^2} W_{pq}(X)
+\frac{1}{Z-X}\left(\frac{1}{k}(W^2)_{pq}(X)-DW_{pq}(X)
\right)+ \nonumber\\
&& -\frac{\eta-\theta}{Z-X}D^2W_{pq}(X)
\\
\ncon{U(Z)U(X)} &=& \frac{-kn}{n-1} \frac{1}{Z-X} \hs{42}
\ncon{U(Z)N_{pq}(X)} \,=\, 0 \\
\ncon{N_{pq}(Z)N_{rs}(X)} &=& \frac{\eta-\theta}{Z-X} \left\{
\!{\mbox{\ }}^{\mbox{\ }}\!
\delta_{qr}N_{ps}(X) -\delta_{sp}N_{rq}(X)\right\} +\frac{k}{Z-X}
\left(\delta_{pr}\delta_{qs}-\frac{1}{n}\delta_{rs}\delta_{pq} \right)
\label{eq:f}
\ena
with the convention that the indices $i,j,l,m$ run from 1 to $n+1$,
while the indices $p,q,r,s$ run from 1 to $n$ (and summation over
repeated indices). I have also introduced:
\be
W_{pq}(X)= N_{pq}(X)-\delta_{pq}\frac{n-1}{n}U(X)
\ee

As announced in \cite{nous}, we get an $N=2$ (linear) superconformal
algebra $\{T(Z),G_0(Z)\}$. This algebra commutes with the super
KM algebra $Sl(n)\oplus U(1)$ built on $\{U(Z),N_{pq}(Z)\}$.
When adding the superstress
energy tensor of this super KM algebra, all the fields become
primary under the stress-energy tensor $T_{tot}$:
\bea
T_{{tot}}(Z) &=& \frac{1}{k^2}T(Z)+T_{Sl(n)}(Z)+T_{U(1)}(Z) \\
T_{Sl(n)}(Z) &=& \frac{1}{3k^2} Str(N^3) +\frac{1}{2k}Str(NDN) \label{Tsln}\\
T_{U(1)}(Z) &=& -\frac{1}{2k}\frac{n-1}{n}UDU \label{Tu1}\\
\ncon{T_{{tot}}(Z)U(X)} &=& \half\frac{\eta-\theta}{(Z-X)^2}U(X)
+\frac{1/2}{Z-X}DU(X) +\frac{\eta-\theta}{Z-X} D^2U(X) \\
\ncon{T_{{tot}}(Z)N_{pq}(X)} &=& \half\frac{\eta-\theta}{(Z-X)^2}N_{pq}(X)
+\frac{1/2}{Z-X}DN_{pq}(X) +\frac{\eta-\theta}{Z-X} D^2N_{pq}(X) \\
\ncon{T_{{tot}}(Z)G_0(X)} &=& \frac{\eta-\theta}{(Z-X)^2}G_0(X)
+\frac{1/2}{Z-X}DG_0(X) +\frac{\eta-\theta}{Z-X} D^2G_0(X) \\
\ncon{T_{{tot}}(Z)G_p(X)} &=& \frac{\eta-\theta}{(Z-X)^2}G_p(X)
+\frac{1/2}{Z-X}DG_p(X) +\frac{\eta-\theta}{Z-X} D^2G_p(X) \\
\ncon{T_{{tot}}(Z)T_{{tot}}(X)} &=&
\frac{3}{2}\frac{\eta-\theta}{(Z-X)^2}T_{{tot}}(X)
+\frac{1/2}{Z-X}DT_{{tot}}(X)
+\frac{\eta-\theta}{Z-X} D^2T_{{tot}}(X)+ \nonumber\\
&& +\frac{k}{(Z-X)^3}
\ena
However, I failed in finding something to add to $G_0(Z)$ for keeping the
$N=2$ linear superconformal algebra. It is some how surprising, since
the superfields $G_p$ and $\bar{G_p}$ have the right $U(1)$
hypercharge to be $N=2$ superfields\footnote{I recall that a $N=1$
superfield $\Phi$ of superspin $s$ can be seen as a $N=2$ superfield of
$U(1)$ hypercharge $\pm 2s$ if it transforms as
$G_0(Z)\Phi(X)=\pm\left\{ 2s\frac{\Phi(X)}{Z-X}+
\frac{\eta-\theta}{Z-X}D\Phi(X)\right\}$}.
With respect to
$T_{{tot}}(Z)$, the $W$ algebra seems to be $N=1$ but
not $N=2$ superconformal.

\indent

Now, from the above superOPEs (between superfields), one can deduce
the OPEs (between fields), and, by
comparison with the OPEs obtained in \cite{JOK}, one is able to
give the combination to choose to get a factorized $W$ algebra.
Before doing it, I first compute the free
superfield realization of the \cw\ algebra.

\subsection{Free superfields realization}

\indent

We start from $J_{diag.}$ and make a finite (residual) gauge
transformation to get $J_{DS}$:
\be
J_{diag.} \rightarrow \hat{L}J_{diag.}L^{-1}+k(DL)L^{-1}\equiv J_{DS}
\mb{with} L\in\cg_+
\ee
Demanding the transformed current to be in the DS-gauge fixes all the
(super)parameters of $L$ in terms of the free superfields of
$J_{diag.}$. Putting these expressions into the transformed current
gives the free superfields realization of the \cw\ algebra. I find:
\beano
T(Z) &=& -(M^3)_{n+1,n+1} +(M^2)_{n+1,n+1}M_{n+1,n+1}
+\frac{k}{2}\left( \!{\mbox{\ }}^{\mbox{\ }}\!
(DM)M +MDM \right)_{n+1,n+1}+
\nonumber\\
&& +\left(
-\sqrt{\frac{k}{2}}(M^2)_{n+1,n+1} +2k\sqrt{\frac{n-1}{n+1}}
\Phi_+M_{n+1,n+1} \right)\Phi_-
+k\frac{k}{2}\frac{n-1}{n+1}\Phi_+D\Phi_+ +
\nonumber \\
&& +k\sqrt{\frac{k}{2}\frac{n-1}{n+1}} \left( M_{n+1,n+1} D\Phi_+
+\Phi_+ DM_{n+1,n+1} \right)
+k^2\left( -\frac{1}{2}\Phi_-D\Phi_- -\sqrt{\frac{k}{2}}D^2\Phi_- \right) \\
G_0(Z) &=& \half (M^2)_{n+1,n+1} +\sqrt{\frac{k}{2}}\left( -M_{n+1,n+1}
+\sqrt{\frac{k}{2}\frac{n-1}{n+1}}\Phi_+ \right)\Phi_- +
\nonumber\\
&& +kD\left( -M_{n+1,n+1}
+\sqrt{\frac{k}{2}\frac{n-1}{n+1}}\Phi_+ \right) \\
G_p(Z) &=& M_{p,q}M_{q,n+1} -\sqrt{\frac{k}{2}}M_{p,n+1}\left(
\Phi_- -\sqrt{\frac{k}{2}\frac{n-1}{n+1}}\Phi_+ \right)
-kD M_{p,n+1} \\
\bar{G}_p(Z) &=& - M_{n+1,q}M_{q,p} -\sqrt{\frac{k}{2}}\left(
\Phi_- +\sqrt{\frac{k}{2}\frac{n-1}{n+1}}\Phi_+ \right)M_{n+1,p}
+kD M_{n+1,p}\\
N_{pq}(Z) &=& M_{pq} +\frac{1}{n}\delta_{pq}M_{n+1,n+1} \hs{24}
U(Z) \,=\, -M_{n+1,n+1} +\sqrt{2k\frac{n^2}{n^2-1}}\Phi_+ \\
&\mbox{with}& \Phi_+=\sqrt{\frac{1}{2k}\frac{n-1}{n+1}}(\Phi_1+\Phi_2)
\mb{and} \Phi_-= \sqrt{\frac{1}{2k}}(\Phi_1-\Phi_2)
\enano
The total stress energy tensor is given by
$T_{tot}(Z)=\frac{1}{k^2}T(Z)+T_{Sl(n)}(Z)+T_{U(1)}(Z)$
with $T_{Sl(n)}(Z)$ and $T_{U(1)}(Z)$ given in (\ref{Tsln}-\ref{Tu1}).
It writes
\bea
T_{tot}(Z) &=& -\half \Phi_+D\Phi_+ -\half \Phi_-D\Phi_-
-\sqrt{\frac{k}{2}}D^2\Phi_- +T_{Sl(n+1)}(Z) \\
T_{Sl(n+1)}(Z) &=& \frac{1}{3k^2}Tr(M^3) +\frac{1}{2k}Tr(MDM)
\ena
{}From the form of $T_{tot}$, one sees that $\Phi_\pm$ are two free
superfields of superspin $\half$, and that $\Phi_+$ is primary,
whereas $\Phi_-$ is not. This is in agreement with the last remark of
section \ref{osp12}, since $\Phi_-$ is carried by $H$.

\indent

Using the superOPEs
\bea
\ncon{\Phi_\pm(Z)\Phi_\pm(X)} &=& \frac{-1}{Z-X} \hs{24}
\ncon{\Phi_+(Z)\Phi_-(X)} \,=\, 0 \\
\ncon{M_{ij}(Z)M_{lm}(X)} &=&
\frac{\eta-\theta}{Z-X} \left\{ \!{\mbox{\ }}^{\mbox{\ }}\!
\delta_{jl}M_{im}^{\ }(X) -\delta_{im}M_{li}(X)\right\} +\frac{k}{Z-X}
\left(\delta_{jl}\delta_{im}-\frac{\delta_{ij}\delta_{lm}}{n+1} \right)
\ena
and the above free superfields realization, one can compute the OPEs
between the \cw\ superfields, and check that we recover the superOPEs
(\ref{eq:i}-\ref{eq:f}).

\subsection{Factorisation of the fermions}

\indent

To factorize out the fermions in the \cw\ algebra, we have to use
the fields expression (we want to fix the gauge for the supersymmetry). Let
us first introduced some notation for the component fields of the
\cw-generators:
\be
\begin{array}{ll}
T(Z) \,=\, \gamma(z)+\theta t(z) & G_0(Z) \,=\, g_0(z)+\theta b_0(z)
\\
G_p(Z) \,=\, g_p(z)+\theta b_p(z) & \bar{G}_p(Z) \,=\,
\bar{g}_p(z)+\theta \bar{b}_p(z)  \\
N_{pq}(Z) \,=\, \lambda_{pq}(z)+\theta \nn_{pq}(z) & U(Z) \,=\,
\eps(z)+\theta u(z) \\
& \\
\Phi_\pm(Z) \,=\, \psi_\pm(z)+\theta\prt\var_\pm(z) &\\
M_{ij}(Z) \,=\, \mu_{ij}(z)+\theta m_{ij}(z) &
\end{array}
\ee
The $n^2$ fermions that we want to factorize are the supersymmetric
partners of the super KM $Sl(n)\oplus U(1)$: $\eps(z)$ and
$\lambda_{pq}(z)$ ($p,q=1,..,n$ with $tr\lambda=0$). I will denote
with a "$\0{\ }$" the quantities built from the $W$ generators
that (anti)commute with these fermions, and say that they are
"free from fermions". For instance, as $N_{pq}(Z)$ and $U(Z)$ anticommute
and because $U(Z)$ is a $U(1)$ current, $u(z)$ is free from fermions:
$\0{u}(z)=u(z)$. On the contrary, it is (of course) impossible to
construct with $\eps(z)$ something free from fermions:
$\0{\eps}=0$.

\subsubsection{Computation of the factorized algebra}

\indent

Let us start with the stress-energy tensor. $t(z)$ is the stress
energy tensor for the whole \cw\ algebra, and we want to keep from it
the part that commutes with the fermions. For such a purpose, one just
subtracts the stress energy tensor of the fermions:
\be
\0{t}(z)= t(z) +\frac{1}{2k}tr(\lambda(z)\prt\lambda(z))
-\frac{n-1}{2nk}\eps(z)\prt\eps(z)
\ee
We can apply the same procedure to the $Sl(n)$ KM algebra $\nn_{pq}(z)$:
we just have to subtract the fermionic realization of this algebra to
get a KM algebra free from fermions:
\be
\0{\nn}_{pq}(z)= \nn_{pq}(z)-\frac{1}{k}\lambda_{pr}(z)\lambda_{rq}(z)
\mb{and} \0{u}(z)=u(z)
\ee
Since we are dealing with classical fields, the level of \0{\nn} is the
same as the one of \nn: $\0{k}=k$.
The remaining spin 1 fields $b_0(z)$, $b_p(z)$ and $\bar{b}_p(z)$
do not required any
work, since their OPEs show that they already commute with the
fermions:
\be
\0{b_0}(z)=b_0(z) \mb{;} \0{b}_p(z)=b_p(z)
\mb{and} \0{\bar{b}}_p(z)=\bar{b}_p(z)
\ee
Note that although $\0{\nn}_{pq}(z)$, $\0{b}_r(z)$,
$\0{\bar{b}}_s(z)$,
$\0{b}_0(z)$ and $\0{u}(z)$
form an $Sl(n+1)\oplus U(1)$ KM algebra, they are not the usual basis
for this algebra, since the $Sl(n)$ part has been singled out. I give
the connection with the usual basis in the next section.

Let us look at the spin 3/2 fields. $g_0(z)$ is already free from
fermions, as it can be seen on the superOPEs between $G_0(Z)$ and
$N_{pq}(Z)$, $U(Z)$. $\gamma(z)$ being the generator of
the linear supersymmetry, we deduce that the part we have to
subtract is the just the supersymmetry generator of an $Sl(n)\oplus
U(1)$ superKM algebra.
\be
\0{\gamma}(z)=\gamma(z) +\frac{n-1}{nk}\eps(z) u(z)
-\frac{2}{3k^2}tr(\lambda^3)(z) -\frac{1}{k}tr(\lambda(z) \nn(z))
\mb{and} \0{g_0}(z)=g_0(z)
\ee
The treatment of $\gamma(z)$ differs from the one of $t(z)$
because the former requires the whole $Sl(n)\oplus U(1)$ algebra, while the
latter involves only the fermions. This means that we cannot make the
factorization at the superfield level, as it should be clear from the
section \ref{osp12}.
Finally, looking at the superOPEs of $G_p(Z)$ and $\bar{G}_p(Z)$ with
$N_{pq}(Z)$ and $U(Z)$, it is quite easy to deduce:
\bea
\0{g}_p(z) &=& g_p(z)-\frac{1}{k}\lambda_{pq}(z)b_q(z)
+\frac{n-1}{2nk}\eps(z)b_p(z)\\
\0{\bar{g}}_p(z) &=& \bar{g}_p(z) +\frac{1}{k}\bar{b}_q(z)\lambda_{qp}(z)
-\frac{n-1}{2nk}\bar{b}_p(z)\eps(z)
\ena

\subsubsection{Comparison with the $Sl(2)$ $W$ algebra}

\indent

Let us first compare our free fields with the ones of \cite{JOK}. The
spin contents in the diagonal gauge is the same for the two
approaches. However, the fermions in \cite{JOK} do not belongs to
$\ca_0$ (see section \ref{sl2.gaug}). Moreover, the fermions $\mu_{ij}(z)$
form a representation of the spin 1 bosons $m_{ij}(z)$,
whereas in \cite{JOK} the bosons commute with the fermions.
Thus, even in the diagonal gauge,
we have to make some combination to recover the fields of \cite{JOK}.
In this gauge, the task is easy. Using a superscript "\jok" (for
Jens-Ole and Katsuchi) to denote the fields of \cite{JOK}, we get:
\be
\begin{array}{ll}
\chi_{n+1}^{\jok}(z)=
\frab{1}{\sqrt{k}}\left(\mu_{n+1,n+1}+\sqrt{\frab{k}{2}}\psi_-
-\sqrt{\frab{k}{2}\frab{n-1}{n+1}}\psi_+\right)
&\chi_{p}^{\jok}(z)= \frab{1}{\sqrt{k}}\mu_{p,n+1}(z) \\
\bar{\chi}_{n+1}^{\jok}(z)=
\frab{1}{\sqrt{k}}\left(\mu_{n+1,n+1}-\sqrt{\frab{k}{2}}\psi_-
-\sqrt{\frab{k}{2}\frab{n-1}{n+1}}\psi_+\right)
&\bar{\chi}_{p}^{\jok}(z)= \frab{1}{\sqrt{k}}\mu_{n+1,p}(z) \\
\hat{J}^{\jok}_{ij}(z)= m_{ij}(z) -\frab{1}{k}\mu_{il}(z)\mu_{lj}(z) &
\end{array}
\label{PIM}
\ee

For the $W$ generators, we first have to find the usual basis for the
$Sl(n+1)\oplus U(1)$ KM algebra. The procedure is more or less the
inverse of the one used to singled out the $Sl(n)$ subalgebra in
$Sl(n+1)$. We use the $b$'s to complete the algebra, construct a
$U(1)$ part that commutes with them, and relax the traceless condition
for the $Sl(n)$ subalgebra. After some calculations, one finds:
\be
\begin{array}{ll}
J_{p,n+1}(z)= -\frab{1}{k}\0{\bar{b}}_p(z)
& J_{p,n+1}(z)= \frab{1}{k}\0{b}_p(z) \\
J_{n+1,n+1}(z)= \frab{2n}{k(n+1)}\0{b}_0(z)-\frab{n-1}{n+1}\0{u}(z)\ \
& J_{pq}(z)= \0{\nn}_{pq}(z)-\frab{1}{n}\delta_{pq}J_{n+1,n+1}(z) \\
v(z)= 2\0{u}(z)-\frab{2}{k}\0{b}_0(z) &
\end{array}
\ee
Using the \jok-fields introduced in (\ref{PIM}), one recover the
expressions given in \cite{JOK} with:
\be
\begin{array}{ll}
t^{\jok}(z)=\0{t}(z) & \\
2G_{n+1}^{\jok}(z)=
\sqrt{k}\0{\gamma}(z)-\frab{2}{\sqrt{k}}\0{g}_0(z)\ \
&G^{\jok}_p(z)= -\frab{1}{\sqrt{k}}\0{g}_p(z) \\
2\bar{G}_{n+1}^{\jok}(z)= \sqrt{k}\0{\gamma}(z)
+\frab{2}{\sqrt{k}}\0{g}_0(z)
&\bar{G}^{\jok}_p(z)= -\frab{1}{\sqrt{k}}\0{\bar{g}}_p(z) \\
J^{\jok}_{ij}(z)=J_{ij}(z) & u^{\jok}(z)=v(z)
\end{array}
\ee
which end the proof of the equality between the factorized \cw-algebra and
the $W$-algebra obtained from $Sl(2)$ reduction.

\indent

One may wonder if it is possible to add more fermions to the
\jok-algebra, so that the other non-linear supersymmetry generators $G_p$
and $\bar{G}_p$ become linear. Let me first remark that the $W$
algebras we get have central extension terms, so that the
linearization will be possible only up to $N=4$ linear
supersymmetries \cite{centr}. Moreover, the $OSp(1|2)$ reduction
give rise to $N=1,2,3,4$ linear supersymmetry, depending on the choice
of the $OSp(1|2)$ under consideration (see last section of \cite{nous}
for more details). Thus, it seems clear that the factorization of spin
1/2 fields between $OSp(1|2)$ and $Sl(2)$ reduction will directly
provide the maximal number of linear supersymmetries. However, the
possibility for these linear supersymmetries to be associated to the
total stress energy tensor is still an open question (see end of
section \ref{supope}).

\sect{Conclusion}

\indent

In this paper, I have considered the Toda models constructed from
the Hamiltonian reduction of a WZW theory built on
a Lie superalgebra. I have
shown that the action S obtained by the reduction with respect to an
$Sl(2)$ subalgebra can be set manifestly supersymmetric if and only
if the $Sl(2)$ algebra is the bosonic part of an $OSp(1|2)$
superalgebra. The two actions S and \cs\ obtained from $Sl(2)$ and
$OSp(1|2)$ reductions are related in the following way: starting from
\cs, one first fixes the gauge for the supersymmetry and
get an action $\cs_0$.
Then, there is a choice of (KM-)gauge where $\cs_0$ and $S$ differs only
by kinetic term of spin 1/2 fields (bosons and fermions).
As a consequence, the $W$ algebra which is a symmetry of S can be
deduced from the super \cw\ algebra, symmetry of \cs, by factorizing
out the spin 1/2 fields. This relation has been explicitly carried
out for the quadratic
superconformal algebra built on $Sl(n+1|2)$.

\indent

A lot of problems are still open on this subject. First, one can
wonder whether this kind of relation is still valid at the quantum
level. As the spin contents of the $W$ (and \cw) algebras are not
changed, one can already notice that these quantum algebras will
differ only by spin 1/2 fields. However, a general treatment on the
quantum actions is still to be done. The detailed study of
"quasi-Abelian" $W$ algebras has to be achieved both for classical and
quantum case.
One may also wonder if some $W$
algebras can be related to some others by a kind of factorization of
higher spins: such studies are in progress \cite{pif}. Of course,
the factorization of the bosonic spin 1/2 fields have to be considered
in details. In particular, a proof of the factorization apart from
the Toda action environment is needed. Note also that, even for spin
1/2 fermions, there exists no systematic way for doing the factorization:
this problem will be solve in \cite{pif}. Finally, the total number of
linear supersymmetries closing on the total stress energy tensor (or
another weaker stress energy tensor) has to be studied.

\newpage

{\Large{\bf Acknowledgements}}

\indent

I would like to thank K. Ito, J.O. Madsen and J.L. Petersen for a lot
of fruitful discussions, and F. Delduc and P. Sorba for reading this
manuscript and many advice.

\end{document}